\documentclass[pra,twocolumn,superscriptaddress,aps,floatfix]{revtex4-2}

\PassOptionsToPackage{hyphens}{url}

\usepackage[T1]{fontenc}
\usepackage[utf8]{inputenc}

\usepackage{hyperref}
\usepackage{amsmath}
\usepackage{bm}
\usepackage{graphicx}
\usepackage{array}
\usepackage{color}
\usepackage{amsxtra} 
\usepackage{amstext}
\usepackage{amssymb}
\usepackage{latexsym}
\usepackage{dsfont}
\usepackage{braket}
\usepackage[caption=false]{subfig}
\usepackage[makeroom]{cancel}
\usepackage{lineno}
\usepackage{physics}
\usepackage{orcidlink}

\usepackage{dcolumn}
\usepackage{bm}
\usepackage{bbm}
\usepackage{braket}
\usepackage{mathrsfs}
\usepackage{amstext}
\usepackage{euscript}

\usepackage{txfonts}

\usepackage{xcolor}
\definecolor{dgreen}{rgb}{0.0,0.6,0.0}
\definecolor{pink}{rgb}{1,0,0.9}

\usepackage[normalem]{ulem}
\usepackage{algorithm}
\usepackage{algpseudocode}

\newcommand{\ms}[1]{\mbox{\scriptsize #1}}

\newcommand{\cref}[1]{Ref.\,\cite{#1}}

\begin{document}

\title{Symplectic split-operator method for the time-dependent unitary Tavis-Cummings model}

\author{Roman Ovsiannikov\orcidlink{0009-0003-9558-4132}}
\email{roman.ovsiannikov@kipt.kharkov.ua}
\affiliation{Akhiezer Institute for Theoretical Physics, NSC KIPT, Akademichna 1, 61108 Kharkiv, Ukraine}

\author{Kurt Jacobs\orcidlink{0000-0003-0828-6421}}
\email{dr.kurt.jacobs@gmail.com}
\affiliation{Advanced Photonic and Electronic Sciences Division, U.S. Army DEVCOM Army Research Laboratory, Adelphi, Maryland 20783, USA} 
\affiliation{Department of Physics, University of Massachusetts at Boston, Boston, Massachusetts 02125, USA}

\author{Andrii G. Sotnikov\orcidlink{0000-0002-3632-4790}}
\email{a\_sotnikov@kipt.kharkov.ua}
\affiliation{Akhiezer Institute for Theoretical Physics, NSC KIPT, Akademichna 1, 61108 Kharkiv, Ukraine}
\affiliation{Education and Research Institute ``School of Physics and Technology'', Karazin Kharkiv National University, Svobody Square 4, 61022 Kharkiv, Ukraine}

\author{Denys I. Bondar\orcidlink{0000-0002-3626-4804}}\email{dbondar@tulane.edu}
\affiliation{Department of Physics and Engineering Physics, Tulane University,  New Orleans, Louisiana 70118, USA}

\date{\today}
\begin{abstract} 

We present a fast, memory-efficient, unitarity-preserving numerical method beyond the rotating-wave approximation for the closed Tavis-Cummings model in which a multilevel spin system interacts with a cavity mode.  
This model can describe the interaction of an ensemble of spins with a cavity mode in which the spin frequency and other parameters are time-dependent. The method exploits the fact that, while the Tavis-Cummings model is not tri-diagonal, it can be brought into tri-diagonal form by a change of basis that can be implemented purely by re-indexing (permuting basis elements), which is a fast operation. By truncating the Fock basis of the cavity mode, the computational complexity of the method is linear in the total dimension of the coupled system, both in time and memory. The method can be employed to simulate any closed quantum system whose Hamiltonian terms can be brought into tri-diagonal form. 
%When solving the evolution problem for the Tavis-Cummings system with a $2N+1$ level atom, without damping, in the non-RW approximation and with time-dependent NV center frequencies, we obtained undefined and non-unitary behavior of the solver in the QuTip package for certain values of the system parameters. Therefore, we developed our own algorithm that allows for unitary evolution with an algorithm complexity of $\sim \mathcal{O}(D)$ in time and $\sim \mathcal{O}(D)$ in memory, where $D$ is the size of the full state vector. Using this algorithm, it is possible to calculate the evolution of any undamped systems whose Hamiltonians can be represented as a set of tridiagonal matrices.

\end{abstract}
\maketitle

\section{Introduction}

Accurate simulation of driven quantum dynamics is essential in cavity quantum electrodynamics, quantum control, and the study of nonclassical-state generation. In these problems, one often needs long-time propagation for explicitly time-dependent Hamiltonians, and the numerical method must remain both efficient and physically faithful. For closed systems, this means, in particular, preserving the unitary structure of the evolution as the Hilbert-space dimension grows \cite{Johansson2012,Johansson2013,Blanes2024,Schneider2025TDSE,Trotter1959,Suzuki1976}. For structure-preserving propagation, symmetric split-operator and symplectic schemes provide a natural starting point, from Strang's classical second-order composition to later quantum-dynamical and symplectic formulations tailored to the time-dependent Schr\"odinger equation \cite{Strang1968,FeitFleckSteiger1982,GrayManolopoulos1996}.  Recent studies have further shown that symplectic split-operator propagators can be constructed efficiently for the Bose--Hubbard Hamiltonian, which is represented as a sum of tridiagonalized many-body operators, extending the scope of structure-preserving simulation to broader classes of quantum models \cite{BondarSteuernagel2026}.

A physically relevant class of the outlined problems is provided by hybrid cavity-spin systems described by Jaynes--Cummings- and Tavis--Cummings-type models \cite{Jaynes1963,Tavis1968,TavisCummings1969,ShoreKnight1993}. Nitrogen-vacancy (NV) centers in diamond are especially attractive in this context because they combine optical addressability, long coherence times, and compatibility with microwave resonators \cite{Doherty2013,Amsuss2011,Kubo2010,Breeze2017}. From the viewpoint of hybrid quantum platforms, collective spin-photon coupling in cavities was identified early as a promising route for spin-ensemble cavity QED \cite{Imamoglu2009}, and high-cooperativity spin-cavity coupling was subsequently demonstrated experimentally~\cite{Schuster2010}. Recent work has further expanded the scope of cavity-NV platforms, including cavity-enhanced magnetometry~\cite{Eisenach2021}, room-temperature cavity-QED effects~\cite{Zhang2022PRL}, microwave-mode cooling~\cite{Ng21,Fahey23,Zhang2022npj}, superradiant masing~\cite{Wu2022Maser,Kersten2026}, spin-refrigerated sensing~\cite{Wang2024Sensor}, radiowave detection~\cite{Hermann2024}, broadband magnetic spectroscopy~\cite{Yin2025}, cavity-enhanced gyroscopy~\cite{Wang2025Gyro}, ultra-low-power frequency combs~\cite{Wang26comb}, and enhanced sensing exploiting a bistable phase transition~\cite{Wang26BP}. In addition, recent theoretical studies of parametrically driven hybrid cavity-NV systems have shown that modulation of the spin frequency can generate amplification and squeezing, and that optimized driving protocols can further enhance these effects \cite{Ovsiannikov2026Hybrid,Ovsiannikov2026Optimal}.

At the same time, direct simulation of finite driven spin-cavity systems remains nontrivial. Although the weak-excitation regime can often be treated by means of the Holstein--Primakoff transformation \cite{Holstein1940}, this approximation does not remove the need for explicit finite-dimensional propagation when counter-rotating terms are retained or when one studies long-time dynamics under strong time-dependent driving. General-purpose solvers are highly versatile, but they do not automatically exploit special sparse structure in a given Hamiltonian and may therefore become unnecessarily expensive for such tasks \cite{Johansson2012,Johansson2013,Schneider2025TDSE}. Recently, structure-exploiting algorithms have been developed for the Tavis–Cummings model that leverage the conservation of the total excitation number and collective angular momentum to enable rapid simulation of thermal properties in mesoscopic ensembles~\cite{Gunderman2025}; here we instead target the full time-dependent unitary dynamics beyond the rotating-wave approximation.

In this study, we consider an undamped driven Tavis--Cummings-type model motivated by hybrid cavity-NV platforms and exploit the fact that its Hamiltonian can be decomposed into a diagonal time-dependent term and two time-independent terms that are tridiagonal in suitable basis orderings. This makes it possible to construct a symmetric Trotter--Suzuki propagation scheme in which the diagonal part is applied exactly, while the tridiagonal parts are treated either by blockwise exponentiation~\cite{Trotter1959,Suzuki1976, bondar_observables_2025} or by a Cayley-type Pad\'{e} approximantion~\cite{Blanes2024}. In the latter case, each time step reduces to the solution of tridiagonal linear systems, yielding a propagation cost that scales linearly with the dimension of the full state vector.

We thus develop a unified propagation algorithm for closed quantum systems whose Hamiltonians admit such a diagonal-plus-tridiagonal decomposition. As a specific example, we apply the method to a finite driven cavity-spin model relevant to hybrid cavity-NV physics. The proposed approach provides a structure-preserving and computationally efficient alternative to generic propagation methods while retaining the physical transparency needed for the analysis of resonantly driven dynamics. 

The paper is organized as follows. In Sec.~\ref{sec:system}, we introduce the model and the Hamiltonian decomposition, then describe the two realizations of the propagation scheme, and finally compare their performance. In Sec.~\ref{sec:conclusion}, we summarize the main advantages and discuss the further applicability of the method.

\section{Description of the algorithm}\label{sec:system}

As an example of a standard method, we first use the QuTiP package~\cite{Johansson2012,Johansson2013,Lambert2025, Ovsiannikov2026Hybrid} to solve the Schr\"odinger equation directly. The Hamiltonian for the system is
\begin{equation}\label{eq1}
    \hat{H} = \omega_{\ms{c}} a^\dagger a + [\omega_{\ms{s}} + \Lambda  \sin(\omega t)]J_z + g (a + a^\dagger)(J_- + J_+) , 
\end{equation}
where $\omega_{\ms{c}}$ is the cavity-mode frequency, $\omega_{\ms{s}}$ is the frequency of the NV centers in the absence of modulation, and $\Lambda$ and $\omega$ are the amplitude and frequency of the modulation, respectively. The cavity-mode bosonic annihilation operator is $a$ (with the total number of modes $N_{\ms{c}}$); $J_-$ is the lowering (ladder) operator for the spin of the size~$J$ with $J_+ = J_-^\dagger$, and $J_z$ is the operator of the spin projection along $z$ axis. Now we can construct the evolution of the eigenvalues of such a covariance matrix:
\begin{equation}\label{EqCovarianeMatrix}
    C_a =  \langle \mathbf{v}_a \mathbf{v}_a ^{\ms{T}} \rangle - \langle \mathbf{v}_a \rangle \langle\mathbf{v}_a ^{\ms{T}} \rangle,
\end{equation}
where $\mathbf{v}_a^{\ms{T}} = (X_a,  Y_a)$ and $X_a = (a+a^\dagger)/2$ and $Y_a = (a-a^\dagger)/(2i)$ \cite{Ovsiannikov2026Hybrid,Ovsiannikov2026Optimal}. 

It is interesting to analyze the resonance case, where $\omega_c + \omega_s = \omega$~\cite{Ovsiannikov2026Hybrid, Ovsiannikov2026Optimal}. We have already studied this problem in the limit of the large spin $J \rightarrow \infty$, but what if $J$ is finite, and the Holstein--Primakoff transformation is not applicable~\cite{Holstein1940}? In the initial moments of time, before the initial ground state is excited to higher spin levels, the solution for the finite system should coincide with the Holstein--Primakoff limit~\cite{Ovsiannikov2026Hybrid}. In principle, if we look at the time evolution shown in Fig.~\ref{fig:evol} and compare the evolution curves of the eigenvalues of the covariance matrix for the Holstein--Primakoff approximation (blue and orange solid lines) with those from the QuTiP solver (green and red dashed lines), we can see a certain agreement. However, this algorithm is rather slow and run-time grows quadratically with the dimension of the state vector (green line in Fig.~\ref{fig:comp_time}).

Therefore, our main goal is to develop a faster algorithm that solves the problem of time evolution for the given Tavis--Cummings model~\eqref{eq1} and preserves unitarity. Note that the Hamiltonian itself explicitly depends on time.

\subsection{Hamiltonian decomposition}

Let us study the evolution of the initial state in the form $\ket{n} \otimes \ket{J, m}$, where $n=0,...,N_c - 1$, $m = -J,...,J$ \cite{Dicke1954,Tavis1968}. Inspired by~\cite{BondarSteuernagel2026}, we propose the new algorithm utilizing tridiagonal form of the matrices to increase the speed of numerical calculations. To this end, let us represent our Hamiltonian in the following form:
\begin{equation}
    \hat{H} = \hat{H}_0 + \hat{V} + \hat{D}(t),
\end{equation}
where 
\begin{align}
    \hat{H}_0 =  g (a J_+ + a^\dagger J_-), \quad \hat{V} = g (a J_- + a^\dagger J_+), \nonumber\\
    \hat{D}(t) = \omega_{\ms{c}} a^\dagger a +  \Delta(t) J_z
\end{align}
and $\Delta(t) = \omega_{\ms{s}} + \Lambda  \sin(\omega t)$. The key observation is that $\hat{H}_0$ and $\hat{V}$ are both \emph{tridiagonal} in appropriately ordered bases, and $\hat{D}(t)$ is diagonal in any product basis.

\paragraph{Basis for $\hat{H}_0$ and $\hat{D}(t)$.}
The operator $\hat{H}_0$ preserves the quantum number $k = n + m$, since the product $a\,J_+$ maps $\ket{n,m}\to\ket{n-1,m+1}$ and $a^\dagger J_-$ maps $\ket{n,m}\to\ket{n+1,m-1}$. (This same conservation of total excitation has also been exploited — alongside the collective angular momentum — to accelerate thermal-state simulations of the Tavis–Cummings model~\cite{Gunderman2025}.) Ordering the basis states primarily by $n+m+J$ (and breaking ties by $m$) makes $\hat{H}_0$ tridiagonal with the matrix elements
\begin{align}
\label{H_0basis}
    \bra{n-1,m+1}\hat{H}_0\ket{n,m} &= g\sqrt{n\,[J(J+1)-m(m+1)]}, \\
    \bra{n+1,m-1}\hat{H}_0\ket{n,m} &= g\sqrt{(n+1)[J(J+1)-m(m-1)]}.
\end{align}
Matrix elements for operator $\hat{D}(t)$ gives the next formula:
\begin{align}
        \bra{n,m}\hat{D}(t)\ket{n,m} &= \omega_{\mathrm{c}}\, n + \Delta(t) \, m, 
\end{align}

\paragraph{Basis for $\hat{V}$.}
The operator $\hat{V}$ conserves the quantum number $\ell = n - m$, since $a\,J_-$ maps $\ket{n,m}\to\ket{n-1,m-1}$ and $a^\dagger J_+$ maps $\ket{n,m}\to\ket{n+1,m+1}$. Ordering by $n - m$ (with the secondary key $m$) makes $\hat{V}$ tridiagonal with matrix the elements
\begin{align}
    \bra{n-1,m-1}\hat{V}\ket{n,m} &= g\sqrt{n\,[J(J+1)-m(m-1)]}, \\
    \bra{n+1,m+1}\hat{V}\ket{n,m} &= g\sqrt{(n+1)[J(J+1)-m(m+1)]}.
\end{align}

\paragraph{Basis switching by reindexing.}
Crucially, switching between the two orderings does not require a matrix multiplication. Since the same set of basis states $\{\ket{n,m}\}$ is used in both cases, the transformation is a \emph{permutation} of the coefficient vector, which costs $\mathcal{O}(D)$ in time and can be precomputed once during initialization. Below, we denote these permutations as $P_{H_0\to V}$ and $P_{V\to H_0}$.

\subsection{Trotter--Suzuki splitting}
Now, we have almost everything necessary to calculate the evolution for one step $\delta t$. According to the obtained knowledge on how to decompose the Hamiltonian into two tridiagonal matrices (let them be tridiagonal in different bases) and one diagonal matrix, we can write the propagator operator for a step $\delta t$ using the Trotter--Suzuki formula, in its symmetric second-order (Strang) form \cite{Trotter1959,Suzuki1976,Strang1968}:
\begin{align}\label{eq:trotter}
    \hat{\mathcal{T}} & e^{-i\int_t^{t + \delta t} \hat{H}(\tau)d\tau} 
    = e^{-i\delta t\,\hat{H}\left(t+\frac{\delta t}{2}\right)} + \mathcal{O}\left(\delta t^3\right) \notag\\
    &= 
    e^{-i\frac{\delta t}{2}[\hat{D}(t+\frac{\delta t}{2})\,+\hat{H}_0]}\;
    e^{-i\delta t\,\hat{V}}\;
    e^{-i\frac{\delta t}{2}[\hat{D}(t+\frac{\delta t}{2})\,+\hat{H}_0]}  + \mathcal{O}\left(\delta t^3\right) \notag\\
    &= 
    e^{-i\frac{\delta t}{4}\hat{D}(t+\frac{\delta t}{2})}\;
    e^{-i\frac{\delta t}{2}\hat{H}_0}\;
    e^{-i\frac{\delta t}{4}\hat{D}(t+\frac{\delta t}{2})}\;
    e^{-i\delta t\,\hat{V}}\;
    \times\notag\\
    & \qquad e^{-i\frac{\delta t}{4}\hat{D}(t+\frac{\delta t}{2})}  e^{-i\frac{\delta t}{2}\hat{H}_0}\;
    e^{-i\frac{\delta t}{4}\hat{D}(t+\frac{\delta t}{2})} + \mathcal{O}\left(\delta t^3\right).
\end{align}
The diagonal factor $e^{-i\frac{\delta t}{2}\Delta(t)\,J_z}$ is applied exactly as an element-wise phase multiplication. For the tridiagonal exponents, we discuss two realizations below.

\textit{Realization~A: Block-diagonal exponentiation:} Since $\hat{H}_0$ and $\hat{V}$ are tridiagonal with zeros on the off-diagonals at the boundaries between conserved-quantum-number sectors, each matrix decomposes into independent blocks. For each block, to achieve maximum efficiency, we diagonalize using a specialized tridiagonal eigensolver~\footnote{For example, in Python, \texttt{scipy.linalg.eigh\_tridiagonal}.} and compute:
\begin{equation}
    e^{-i\delta t\,H_{\mathrm{block}}} = U\,\mathrm{diag}(e^{-i\delta t\,\lambda_k})\,U^\dagger.
\end{equation}
If the sizes of both subsystems are approximately equal, i.e., $N_c \sim 2J + 1$,  the block sizes scale as $\mathcal{O}(\sqrt{D})$, so the total cost per step is $\mathcal{O}(D^{3/2})$. The exponents are time-independent and can be precomputed.

\textit{Realization~B: Cayley (Crank--Nicolson) rational approximation:} We approximate each tridiagonal exponential using the Cayley transform \cite{CrankNicolson1947}. In the present time-dependent Schr\"odinger setting, this Cayley/Crank--Nicolson step is also closely related to the standard unitary-preserving  propagator~\cite{GoldbergScheySchwartz1967}:
\begin{equation}\label{eq:cayley}
    \ket{\phi} = e^{-i\alpha \hat{H}}\,\ket{\psi} \approx \frac{I - i\frac{\alpha}{2}\hat{H}}{I + i\frac{\alpha}{2}\hat{H}}\,\ket{\psi} + \mathcal{O}(\alpha^3),
\end{equation}
which is equivalent to solving the tridiagonal linear system
\begin{equation}\label{EqImplicitEuler}
    \bigl(I + i\tfrac{\alpha}{2}\hat{H}\bigr)\,\ket{\phi} = \bigl(I - i\tfrac{\alpha}{2}\hat{H}\bigr)\,\ket{\psi}
\end{equation}
for the unknown $\ket{\phi}$. Both the right-hand-side matrix--vector product and the tridiagonal solver (Thomas algorithm \cite{Thomas1949}) cost $\mathcal{O}(D)$, giving a total per-step complexity of the order of $\mathcal{O}(D)$.

Note that for a Hermitian $\hat{H}$, the Cayley transform~\eqref{eq:cayley},
$\left(I - i\frac{\alpha}{2}\hat{H}\right)\left(I + i\frac{\alpha}{2}\hat{H}\right)^{-1}$,
is manifestly unitary; hence, the propagator~\eqref{EqImplicitEuler} from the initial state $\ket{\psi}$ to the wavefunction $\ket{\phi}$ at later times preserves the norm to machine precision.

\subsection{Algorithm}

With the knowledge on how to evaluate one step of time evolution at $\delta t$, let us present the complete algorithm in the form of pseudocode:
\begin{algorithm}[H]
\caption{Split-operator propagation for the driven Tavis--Cummings model}
\label{alg:propagation}
\begin{algorithmic}[1]
\Require System parameters $(\omega_c, g, \Delta(\cdot))$; truncation $(N_c, J)$; time step $\delta t$; number of steps $N_t$; method $\in\{\textsc{Exp},\textsc{Linear}\}$
\Ensure Time-evolved state $\ket{\psi(t_f)}$ and observables

\Statex
\State \textbf{--- Initialization (performed once) ---}
\State Generate basis states $\{(n,m)\}$ with $n=0,\dots,N_c-1$, $m=-J,\dots,J$
\State Sort by $(n{+}m{+}J,\; m)$ $\to$ ordered arrays $(\mathbf{n},\mathbf{m})$ \Comment{$\hat{H_0}$ is tridiagonal}
\State Build tridiagonal $\hat{H_0}$ from Eq.~(5)--(7) using $(\mathbf{n},\mathbf{m})$
\State Sort by $(n{-}m,\; m)$ $\to$ ordered arrays $(\mathbf{n}',\mathbf{m}')$ \Comment{$\hat{V}$ is tridiagonal}
\State Build tridiagonal $\hat{V}$ from Eq.~(8)--(9) using $(\mathbf{n}',\mathbf{m}')$
\State Precompute permutation arrays $P_{H_0\to V}$ and $P_{V\to H_0}$
\If{method $=$ \textsc{Exp}}
    \State Precompute $e^{-i\delta t\,\hat{H_0}/2}$ and $e^{-i\delta t\,\hat{V}}$ via block diagonalization
\EndIf
\State Set initial state $\psi \gets \ket{0}\otimes\ket{J,-J}$ (in $H_0$-ordered basis)

\Statex
\State \textbf{--- Time stepping ---}
\For{$p = 1,\dots,N_t$}
    \State $t \gets t + \delta t/2$ \Comment{Midpoint for $\Delta(t)$}
    \State $\phi_{m, n} \gets e^{-i\,\delta t\,(\Delta(t)\,m + \omega_c\, n)/4}$ for each $n$ and $m$ \Comment{Diagonal phases}

    \Statex
    \State $\psi \gets \phi_{m,n} \odot \psi$ \Comment{$\exp(-i\frac{\delta t}{4}\hat{D}(t))$: element-wise, $\mathcal{O}(D)$}

    \Statex
    \If{method $=$ \textsc{Exp}}
        \State $\psi \gets e^{-i\delta t \hat{H_0}/2}\;\psi$ \Comment{Sparse mat-vec, precomputed}
        \State $\psi \gets \phi_{m, n} \odot \psi$ \Comment{$\exp(-i\frac{\delta t}{4}\hat{D}(t))$ again}
        \State $\psi \gets P_{H_0\to V}[\psi]$ \Comment{Reindex to $V$-basis, $\mathcal{O}(D)$}
        \State $\psi \gets e^{-i\delta t \hat{V}}\;\psi$
        \State $\psi \gets P_{V\to H_0}[\psi]$ \Comment{Reindex back, $\mathcal{O}(D)$}
        \State $\psi \gets \phi_{m, n} \odot \psi$ \Comment{$\exp(-i\frac{\delta t}{4}\hat{D}(t))$ again}
        \State $\psi \gets e^{-i\delta t \hat{H_0}/2}\;\psi$
    \Else \Comment{method $=$ \textsc{Linear}}
        \State Set $\alpha \gets \delta t/4$
        \State Solve $(I + i\alpha \hat{H_0})\,\psi' = (I - i\alpha H_0)\,\psi$ \Comment{Thomas alg., $\mathcal{O}(D)$}
        \State $\psi \gets \phi_{m, n} \odot \psi$ \Comment{$\exp(-i\frac{\delta t}{4}\hat{D}(t))$ again}
        \State $\psi \gets P_{H_0\to V}[\psi']$ \Comment{Reindex, $\mathcal{O}(D)$}
        \State Solve $(I + 2i\alpha \hat{V})\,\psi' = (I - 2i\alpha \hat{V})\,\psi$ \Comment{Thomas alg., $\mathcal{O}(D)$}
        \State $\psi \gets P_{V\to H_0}[\psi']$ \Comment{Reindex, $\mathcal{O}(D)$}
        \State $\psi \gets \phi_{m, n} \odot \psi$ \Comment{$\exp(-i\frac{\delta t}{4}\hat{D}(t))$ again}
        \State Solve $(I + i\alpha \hat{H_0})\,\psi' = (I - i\alpha \hat{H_0})\,\psi$ \Comment{Thomas alg., $\mathcal{O}(D)$}
        \State $\psi \gets \psi'$
    \EndIf

    \Statex
    \State $\psi \gets \phi_{m, n} \odot \psi$ \Comment{$\exp(-i\frac{\delta t}{4}\hat{D}(t))$ again}
    \State $\psi \gets \psi / \|\psi\|$ \Comment{Renormalize}
    \State $t \gets t + \delta t/2$
    \State Compute and store observables from $\psi$
\EndFor
\end{algorithmic}
\end{algorithm}

\subsection{Evolution and complexity comparison}

\begin{figure}[t]
\includegraphics[width=1\columnwidth]{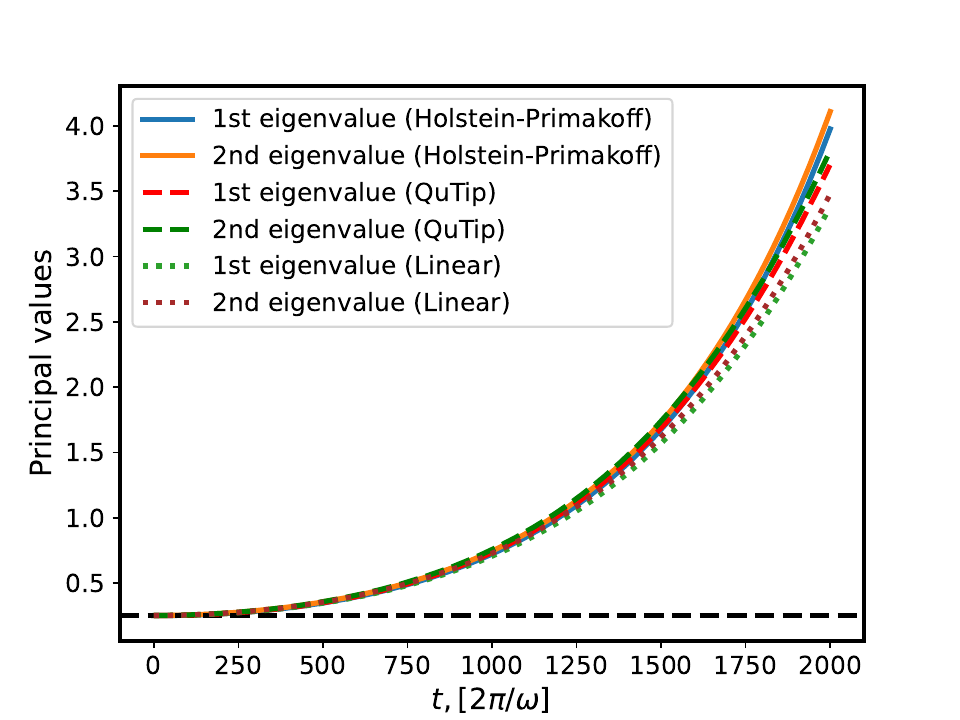} 
    \caption{Evolution of the eigenvalues of the covariance matrix~\eqref{EqCovarianeMatrix} for Holstein--Primakoff approach (orange and blue solid lines), QuTip solver (red and green dashed lines), and our algorithm~\ref{alg:propagation} with \textsc{method=Linear} (light green and brown dotted lines)  with parameters $\omega_c = 2\pi \times 2.4 $~GHz,  $\omega_s = 2\pi \times 3.6 $ GHz, $g = 2 \pi \times 10$ MHz, $\Lambda = 2\pi \times 1 $~GHz and $\omega = 2\pi \times 6 $~GHz. The shape of the system is $N_c = 50$ and $J = 20$.}
    \label{fig:evol}
\end{figure}

To validate the proposed propagation scheme, we compare the time evolution obtained with our algorithm to both the Holstein--Primakoff approximation and direct numerical integration using the QuTiP solver. 
Figure~\ref{fig:evol} shows the evolution of the eigenvalues of the cavity quadrature covariance matrix~\eqref{EqCovarianeMatrix} for a finite spin ensemble. At short times, the finite-system dynamics obtained using the proposed method (light green and brown dotted lines) agrees well with both the Holstein–Primakoff approximation and the direct QuTiP solution. At longer times, deviations from the Holstein–Primakoff limit appear due to finite-size effects, while our method remains consistent with the direct numerical solution.

The developed propagation scheme is based on a second-order symmetric Trotter--Suzuki splitting \cite{Suzuki1976}, so that the local error of one time step is $\mathcal{O}(\delta t^3)$ and the global error is $\mathcal{O}(\delta t^2)$ for a fixed total evolution time. 
The diagonal factor $\Delta(t)J_z$ is applied exactly, while the tridiagonal parts are treated either by direct blockwise exponentiation or by the Cayley rational approximation \cite{CrankNicolson1947}.
The latter replaces exponentiation by the solution of tridiagonal linear systems and is numerically stable and near-unitary for Hermitian Hamiltonians. 
In long runs, small roundoff-induced norm deviations may accumulate, so we explicitly renormalize the state vector after each step. 
In practice, the time step $\delta t$ must be chosen small enough to resolve both the external modulation and the fastest intrinsic oscillation scales.

\begin{figure}
\includegraphics[width=1\columnwidth]{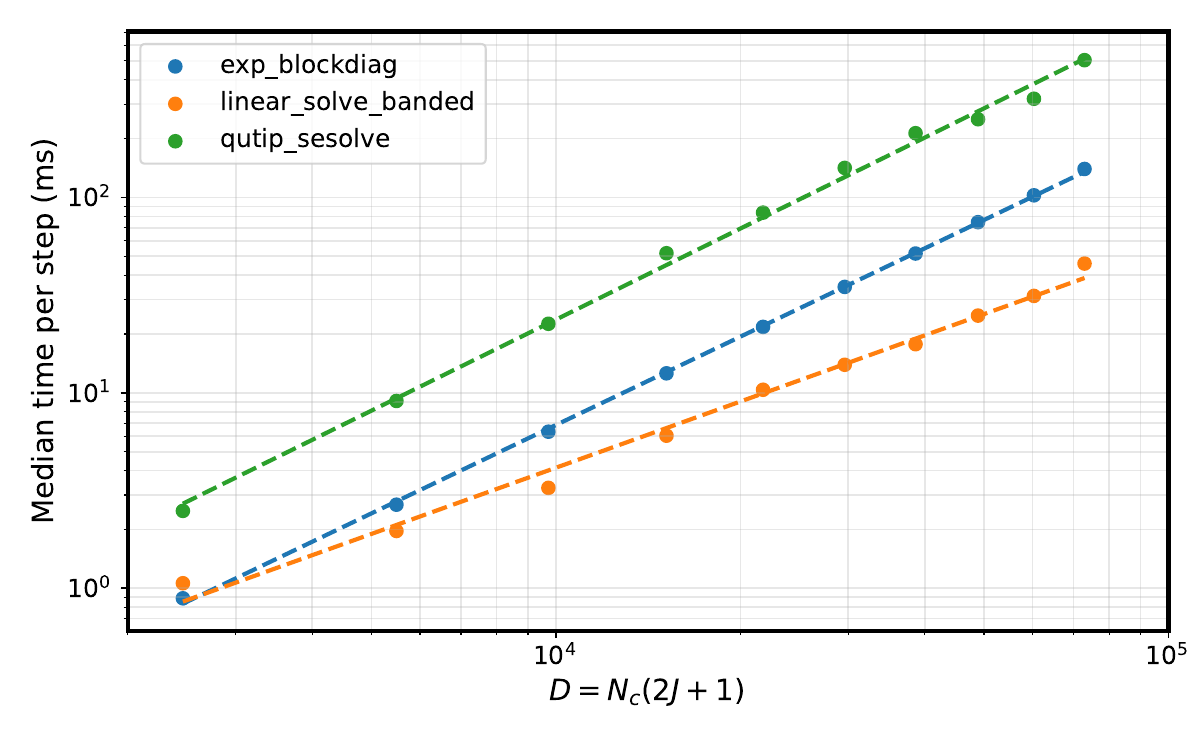} 
    \caption{Time of the one step evolution for Qutip solver (green line), our algorithm ~\ref{alg:propagation} with \textsc{method=Exp} (blue line), and with \textsc{method=Linear} (orange line). For each point the shape of the system is a $N_c = \{60, 90, 120, 150, 180, 210, 240, 270, 300, 330\}$ and $J = \{20, 30, 40, 50, 60, 70, 80, 90, 100, 110\}.$}
    \label{fig:comp_time}
\end{figure}
To assess computational efficiency, we compare the runtime of a single propagation step for three approaches: (i) direct integration using QuTiP, (ii) block-diagonal exponentiation, and (iii) the proposed linear Cayley-based method. Figure~\ref{fig:comp_time} shows the runtime as a function of Hilbert-space dimension. Our method (orange line) exhibits approximately linear scaling with system size, while the block-diagonal method (blue line) shows subquadratic scaling, while compared to this the QuTiP solver has similar scaling but a significantly larger prefactor. These results confirm the computational advantage of the proposed approach for large-scale simulations.

The main limitation of the method is that it is tailored to the Tavis--Cummings Hamiltonian~\eqref{eq1} as it relies on the decomposition of the Hamiltonian into diagonal and tridiagonal parts in suitable basis orderings. If additional terms destroy this sparse structure, the computational advantage is reduced or lost. The present formulation is also intended for closed-system unitary dynamics, extension to open system dynamics can be readily achieved by utilizing low-rank corner space techniques recently developed for solving master equations for large open quantum systems~\cite{le_bris_low-rank_2013, finazzi_corner-space_2015, mccaul_fast_2021, chen_low_2020, donatella_continuous-time_2021}. Other possible extensions include higher-order splittings~\cite{Suzuki1990, bandrauk_higher_1992} and generalization to banded or block-tridiagonal Hamiltonians.

\begin{table}
\centering
\begin{tabular}{lcc}
\hline
Method & Time per step & Memory \\
\hline
QuTiP \texttt{sesolve} & $\mathcal{O}(D^{3/2})$ & $\mathcal{O}(D^{3/2})$ \\
Block-diag.\ exponentiation & $\mathcal{O}(D^{3/2})$ & $\mathcal{O}(D^{3/2})$ \\
Cayley / Thomas algorithm & $\mathcal{O}(D)$ & $\mathcal{O}(D)$ \\
\hline
\end{tabular}
\caption{Asymptotic scaling of the three propagation methods, where $D=N_c(2J+1)$ is the total Hilbert-space dimension of the coupled system.}
\label{tab:complexity}
\end{table}

\section{Conclusion}\label{sec:conclusion}

We have developed Algorithm~\ref{alg:propagation} for the unitary propagation of undamped quantum systems whose Hamiltonians can be represented as a sum of diagonal and tridiagonal parts in suitable basis orderings. As a specific example, we applied the method to a driven finite Tavis--Cummings-type model in the non-rotating-wave-approximation regime with time-dependent modulation of the NV-center frequency.

The proposed scheme combines a symmetric Trotter--Suzuki splitting with two possible realizations of the tridiagonal evolution steps: blockwise exponentiation and a Cayley-type rational approximation. The latter (Algorithm~\ref{alg:propagation} with \textsc{method=Linear}) replaces exponentiation by the solution of tridiagonal linear systems and therefore yields a propagation cost that scales linearly with the dimension of the full state vector.

For the model considered here, the algorithm reproduces the expected short-time behavior of the Holstein--Primakoff limit and provides a clear computational advantage over direct propagation with generic solvers. In particular, the linear version of the method is the most efficient among the approaches studied while maintaining near-unitary evolution over long runs.

\section*{Code availability} 

All code used in this study can be found in~\cite{Codes}. 

\acknowledgments

D.I.B. was supported by DEVCOM Army Research Office (ARO) (grant W911NF-23-1-0288; program manager Dr.~James Joseph). R.O. and A.G.S. acknowledge support by the National Research Foundation of Ukraine, project No.~2023.03/0073. The views and conclusions contained in this document are those of the authors and should not be interpreted as representing the official policies, either expressed or implied, of ARO or the U.S. Government. The U.S. Government is authorized to reproduce and distribute reprints for Government purposes notwithstanding any copyright notation herein.

\bibliography{refs.bib}

\end{document}